\documentclass[fleqn,usenatbib,useAMS]{mnras}

\usepackage{graphicx}	
\usepackage{amsmath}	
\usepackage{amssymb}	
\usepackage{multicol}        
\usepackage{bm}		
\usepackage{pdflscape}	

\usepackage[T1]{fontenc}
\usepackage{ae,aecompl}

\usepackage{newtxtext,newtxmath}

\title[Search for the companion of XO-7b]{Search for the wide-orbit massive companion of XO-7b in the follow-up radial-velocity and transit-timing data: no significant clues}

\author[Z. Garai et al.]{Z. Garai,$^{1,2,3}$\thanks{E-mail: zgarai@ta3.sk; zgarai@gothard.hu}
T. Pribulla,$^{3}$
and R. Kom\v{z}\'{i}k,$^{3}$
\\
$^{1}$HUN-REN-ELTE Exoplanet Research Group, 9700 Szombathely, Szent Imre h. u. 112, Hungary\\
$^{2}$ELTE Gothard Astrophysical Observatory, 9700 Szombathely, Szent Imre h. u. 112, Hungary\\
$^{3}$Astronomical Institute, Slovak Academy of Sciences, 05960 Tatransk\'a Lomnica, Slovakia\\
}

\date{Last updated 2020 June 10; in original form 2013 September 5}

\pubyear{2020}

\begin{document}
\label{firstpage}
\pagerange{\pageref{firstpage}--\pageref{lastpage}}
\maketitle

\begin{abstract}
XO-7b is a hot Jupiter transiting a $V = 10.52$ mag G0V-type star. The planetary system is interesting because the linear slope in the discovery radial-velocity (RV) data indicated a wide-orbit massive companion. In 2020 we started an RV campaign for the system with the main scientific goal to follow-up this linear slope, and to put constraints on the orbital period of the companion. Furthermore, we aimed at refining the system parameters and we wanted to probe transit timing variations (TTVs) of XO-7b in order to search for long-term dynamical signs of the companion of XO-7b in the observed-minus-calculated (O-C) data of mid-transit times. Apart from the discovery RVs, we obtained and analyzed 20 follow-up RV observations and \textit{TESS} photometric data. The previously observed significant linear RV slope was not confirmed with the follow-up RV data, where we detected only a marginal linear slope with the opposite trend. If the announced companion really exists, the most convincing explanation is that both RV datasets were collected near its quadrature position. Based on the RVs we estimated the minimum orbital period, which is $P_\mathrm{orb,min,3} \gtrsim 7900 \pm 1660$ d, and the 'minimum' minimum mass of the companion, which is $(M_3 \sin i)_\mathrm{min} = 16.7 \pm 3.5~\mathrm{M_{Jup}}$. We did not find significant evidence of the companion of XO-7b in the O-C dataset of mid-transit times. We can again conclude that if the announced companion really exists, this is in agreement with previous results that distant companions of exoplanets are only known by RV solutions.  
\end{abstract}




\section{Introduction}
\label{intro}

Nowadays the \textit{Transiting Exoplanet Survey Satellite} (\textit{TESS}) mission \citep{Ricker1, Ricker2}, launched in 2018, is the most active transiting-exoplanet searching program. During its nominal and extended mission duration, up to September 19, 2023, discovered 392 confirmed planets and 6788 planet candidates\footnote{This information is based on the NASA Exoplanet Archive database (see \url{https://exoplanetarchive.ipac.caltech.edu/index.html}).}. In comparison with the formerly active \textit{Kepler} \citep{Borucki1, Borucki2, Borucki3} and \textit{K2} \citep{Howell1} projects it has a big advantage, namely that \textit{TESS} targeted bright stars, brighter than $T$ (\textit{TESS}) = 13 mag. Bright stars are needed to get true masses of the exoplanets via ground-based radial velocity (RV) measurements \citep{Mayor1}. Moreover, \textit{TESS} has one more big advantage in comparison with \textit{Kepler}, i.e. that it is performing a near-all-sky survey in sectors. Every sector has a dimension of $24 \times 96$ deg, and the sectors are monitored for at least 27 days. The photometric precision is dominated by pointing jitter. A 1-hour integration at a $T = 10$ mag star gives about 230 ppm precision, which is sufficient to detect super-Earths. These characteristics make this space telescope not only a powerful tool in the discovery of new exoplanets but also a useful tool for the follow-up of known exoplanets. Apart from the \textit{Characterising Exoplanet Satellite} (\textit{CHEOPS}) space observatory \citep{Benz1}, \textit{TESS} is the most used space-based instrument for these purposes. 

Known exoplanets from ground-based surveys were frequently followed-up using \textit{TESS} data. For example, physical and orbital parameters of WASP-18Ab, WASP-19b, and WASP-77Ab were refined based on the \textit{TESS} data by \citet{Cortes1}, and planetary architecture and stellar variability constraints were improved at WASP-166b by \citet{Doyle1}. Transit time variations (TTVs) were probed with precise \textit{TESS} data, e.g., in the cases of WASP-4b \citep{Bouma1, Baluev1}, XO-6b \citep{Ridden1}, or HATS-18b \citep{Southworth1}. The decaying orbit of WASP-12b was confirmed based on \textit{TESS} observations by \citet{Turner1} and \citet{Wong2}. The orbital decay was probed also in the cases of WASP-18b \citep{Maciejewski1} and WASP-43b \citep{Garai1, Davuodi1} using \textit{TESS} data. 

In this follow-up project, we selected the transiting exoplanet XO-7b, discovered by \citet{Crouzet1} within the framework of the XO project \citep{McCullough1}. The XO project discovered seven transiting exoplanets to date \citep{McCullough2, McCullough3, Burke1, Burke2, Johns-Krull1, Crouzet2, Crouzet1}. Although the number of discoveries is relatively small, XO planets are interesting in many aspects. XO-7b is also interesting, because its large atmospheric scale height makes it well suited to atmospheric characterization, and the linear slope in the discovery RV data indicates a wide-orbit companion with a minimum mass of $4~\mathrm{M_{Jup}}$, which could be a planet, a brown dwarf, or a low-mass star \citep{Crouzet1}. In some cases, the outer companions are known to exhibit their signal in TTVs and RV observations, as well, distant companions of exoplanets are only known by RV solutions, see e.g., \citet{Maciejewski2, NeveuVanMalle1, Dawson1, Knutson1}, or \citet{Hartman1}. 

Motivated by the announced wide-orbit massive companion of XO-7b, in 2020 we started RV observations of the XO-7 system with the main scientific goal to follow-up the linear RV slope discovered by \citet{Crouzet1}, and to put constraints on the orbital period of the companion. Furthermore, we aimed at combining published RVs and the RV data from this work with \textit{TESS} photometric data to derive precise orbital and planetary parameters of the XO-7 system. Basic stellar parameters of XO-7 were obtained from spectra, which were later used to extract the RV data. Finally, based on \textit{TESS} data we probed the TTVs of XO-7b. Although results on TTVs based on \textit{TESS} data were recently published by \citet{Maciejewski3}, who found no additional planet in the XO-7 system down to sub-Neptune-sized planets, we repeated the analysis with a much longer timebase \textit{TESS} data in order to search for long-term dynamical signs of the announced companion of XO-7b in the TTV data. 

The paper is organized as follows. In Sect. \ref{obsdatred} a brief description of observations, the applied spectroscopic instrumentation, and data reduction is given. We summarise the spectral analysis and the derived stellar parameters in Sect. \ref{spectrsynthres}. Data analysis, model fitting, and the results are detailed in Sect \ref{dataanres}. The most interesting result is described and discussed in Sect. \ref{discuss}. We conclude with the results in Sect. \ref{concl}.

\section{Observations and data reduction}
\label{obsdatred}

\subsection{Photometric observations}
\label{tessphotdata}

We used 2-min integrated \textit{TESS} data of XO-7, downloaded from the Mikulski Archive for Space Telescopes\footnote{See \url{https://mast.stsci.edu/portal/Mashup/Clients/Mast/Portal.html}.} (MAST). Up to September 30, 2023, XO-7 was observed with \textit{TESS} during 11 sectors, Nos. 18, 19, 20, 25, 26, 40, 47, 52, 53, 59, and 60. In the first three sectors, however, no 2-min integrations were produced, therefore, we downloaded only the last 8 sectors' data in the form of Pre-search Data Conditioning Simple Aperture Photometry (PDCSAP) flux, which was initially smoothed by the PDCSAP pipeline. This light-curve type is subject to more treatment than the Simple Aperture Photometry (SAP) flux and is specifically intended for detecting exoplanets. The pipeline attempts to remove systematic artifacts while keeping planetary transits intact. This light-curve type is also corrected by the dilution factor. From MAST we obtained $134230$ PDCSAP datapoints with 2-min integration (for more details see Tab. \ref{TESSobslog}). These \textit{TESS} data were first normalized to unity. Then, since \textit{TESS} telescope uses as time-stamps Barycentric \textit{TESS} Julian Date (i.e. $\mathrm{BJD_{TDB}} - 2~457~000.0$), we converted all \textit{TESS} time-stamps to $\mathrm{BJD_{TDB}}$. 

\begin{table}
\centering
\caption{Log of 2-min integrated \textit{TESS} PDCSAP photometric observations of XO-7 used in this work. The table shows the time interval of observations, the number of observed transits, and the number of data points obtained from MAST, sorted by the \textit{TESS} sectors.}
\label{TESSobslog}
\begin{tabular}{cccc}
\hline
Sector & Time interval & Number & Data\\
No. & of observation & of transits & points\\ 
\hline
25 & 2020-05-13 -- 2020-06-08 & 9  & 17~246\\ 
26 & 2020-06-08 -- 2020-07-04 & 8  & 16~942\\
40 & 2021-06-24 -- 2021-07-23 & 10 & 19~611\\
47 & 2021-12-30 -- 2022-01-28 & 8  & 17~311\\
52 & 2022-05-18 -- 2022-06-13 & 8  & 16~749\\
53 & 2022-06-13 -- 2022-07-09 & 8  & 17~305\\
59 & 2022-11-26 -- 2022-12-23 & 9  & 16~476\\ 
60 & 2022-12-23 -- 2023-01-18 & 7  & 12~590\\ 
\hline
Total & -- 					& 67 & 134230\\
\hline
\end{tabular}
\end{table}
 
\subsection{Spectroscopic observations and RV data}
\label{spectroobs}

\begin{table*}
\centering
\caption{Log of spectroscopic observations of XO-7, obtained at the Skalnat\'{e} Pleso Observatory. The table shows the time of the observation, BF RVs and CCF RVs with $\pm 1\sigma$ uncertainties, systematic errors caused by the spectrograph instability ($\sigma_\mathrm{syst}$), the signal-to-noise ratio of the combined spectra at 5500 \AA, calculated as $S/N = \sqrt{(S/N)_1{^2} + (S/N)_2{^2} + (S/N)_3{^2}}$, where $(S/N)_\mathrm{n}$ is the signal-to-noise ratio of individual spectra, and the orbital phase of observation, calculated using the mid-transit time of $T_\mathrm{c} = 2~458~779.58040$ $\mathrm{BJD_{TDB}}$ and the orbital period of $P_\mathrm{orb} = 2.86413296$ d (see Tab. \ref{newparameterstab}).}
\label{spectroobslog}
\begin{tabular}{cccccc}
\hline
Time [$\mathrm{BJD_{TDB}}$] & RV (BF) [$\mathrm{km~s}^{-1}$] & RV (CCF) [$\mathrm{km~s}^{-1}$] & $\sigma_\mathrm{syst}$ [$\mathrm{km~s}^{-1}$] & $S/N$ & Orbital phase\\
\hline
2458928.385194 & $-13.750 \pm 0.047$ & $-13.083 \pm 0.073$ & 0.30 & 21.0 & 0.95456\\   
2459108.342633 & $-13.680 \pm 0.032$ & $-13.177 \pm 0.070$ & 0.09 & 30.5 & 0.78595\\   
2459163.384992 & $-13.790 \pm 0.035$ & $-13.259 \pm 0.152$ & 0.17 & 27.9 & 0.00376\\   
2459164.350772 & $-13.840 \pm 0.031$ & $-13.330 \pm 0.177$ & 0.07 & 32.2 & 0.34096\\   
2459196.222682 & $-13.820 \pm 0.050$ & $-13.188 \pm 0.103$ & 0.35 & 19.9 & 0.46890\\   
2459197.676202 & $-13.850 \pm 0.056$ & $-13.198 \pm 0.154$ & 0.05 & 17.7 & 0.97639\\   
2459226.470352 & $-13.760 \pm 0.036$ & $-13.121 \pm 0.116$ & 0.26 & 27.7 & 0.02975\\   
2459624.500127 & $-13.690 \pm 0.047$ & $-13.124 \pm 0.107$ & 0.03 & 20.9 & 0.00017\\   
2459650.456236 & $-13.810 \pm 0.040$ & $-13.210 \pm 0.067$ & 0.14 & 24.6 & 0.06264\\   
2459667.415096 & $-13.750 \pm 0.063$ & $-13.046 \pm 0.209$ & 0.31 & 15.7 & 0.98375\\   
2459699.515815 & $-13.900 \pm 0.061$ & $-13.390 \pm 0.130$ & 0.10 & 16.2 & 0.19158\\   
2459712.420535 & $-13.720 \pm 0.052$ & $-13.137 \pm 0.072$ & 0.50 & 19.1 & 0.69721\\   
2459718.393255 & $-13.660 \pm 0.050$ & $-13.223 \pm 0.068$ & 0.40 & 19.8 & 0.78256\\   
2460065.433550 & $-13.629 \pm 0.048$ & $-13.062 \pm 0.106$ & 0.10 & 20.8 & 0.95022\\   
2460084.480879 & $-13.607 \pm 0.040$ & $-13.167 \pm 0.099$ & 0.00 & 25.1 & 0.60052\\       
2460099.463299 & $-13.766 \pm 0.044$ & $-12.971 \pm 0.069$ & 0.00 & 22.7 & 0.83157\\       
2460114.483999 & $-13.797 \pm 0.056$ & $-13.161 \pm 0.270$ & 0.10 & 17.6 & 0.07598\\       
2460115.453659 & $-13.813 \pm 0.050$ & $-13.194 \pm 0.065$ & 0.10 & 19.7 & 0.41454\\                        
2460142.428958 & $-13.745 \pm 0.049$ & $-13.061 \pm 0.177$ & 0.20 & 20.4 & 0.83285\\
2460153.455588 & $-13.652 \pm 0.050$ & $-13.088 \pm 0.097$ & 0.00 & 19.9 & 0.68275\\
\hline                                 
\end{tabular}                          
\end{table*}

Several spectroscopic observations were performed at the Skalnat\'{e} Pleso Observatory in the High Tatras (Slovak Republic), using the 1.3 m f/8.36 Astelco Alt-azimuthal Nasmyth-Cassegrain reflecting telescope, equipped with a fiber-fed echelle spectrograph of MUSICOS design \citep{Baudrand1}. Its fiber injection and guiding unit (FIGU) is mounted in the Nasmyth focus of the telescope. The FIGU is connected to the calibration unit (ThAr hollow cathode lamp, tungsten lamp, blue LED) in the control room and to the echelle spectrograph itself in the room below the dome, where the temperature is stable. The spectra were recorded by an Andor iKon-936 DZH $2048 \times 2048$ pixels CCD camera. The spectral range of the instrument is 4250 -- 7375 \AA~ in 56 echelle orders. The maximum resolution of the spectrograph reaches $R \approx 38~000$ around 6000 \AA. The exposure time was 900 s in all cases. Three raw spectra were obtained consecutively during an observing night and, subsequently, combined via \texttt{IRAF} task \texttt{combine} to increase the signal-to-noise ratio ($S/N$) with the assumption that there is no substantial difference (Doppler shift) between them. 

The raw spectra were reduced using \texttt{IRAF} package tasks, \texttt{Linux} shell scripts, and \texttt{FORTRAN} programs similarly, as it was described in \citet{Pribulla1} and in \citet{Garai3}. In the first step, master dark frames were produced. In the second step, the photometric calibration of the frames was done using dark and flat-field frames. Bad pixels were cleaned using a bad pixel mask, and cosmic hits were removed using the program of \citet{Pych1}. Order positions were defined by fitting Chebyshev polynomials to the tungsten lamp and blue LED spectrum. In the following step, scattered light was modeled and subtracted. Aperture spectra were then extracted for both object and ThAr frames, and then the resulting 2D spectra were dispersion-solved. Two-dimensional spectra were finally combined to 1D spectra rebinned to 4250 -- 7375 \AA~ wavelength range with a 0.05 \AA~ step, i.e. about 2 -- 4 times the spectral resolution. 

Spectra were first analyzed using the broadening function (BF) technique, developed by \citet{Rucinski1}, to get RVs. As a template star HD\,222368 ($\iota$ Psc, F7V-type) was used with RV = $5.4~\mathrm{km~s}^{-1}$ \citep{Nordstroem1}. The RVs of XO-7 were measured by fitting a Gaussian function to the extracted BFs. Because the formal RV uncertainties found using the BF approach are hard to quantify and depend on BF smoothing, they were determined from $S/N$ as follows. We first determined RV uncertainties as 1/($S/N$), see \citet{Hatzes1}. Subsequently, we fitted the RV observations with initial uncertainties using the \texttt{RMF} code \citep{Szabo1, Garai1, Garai2}, and from the best fit we obtained reduced $\chi^{2}$ ($\chi^{2}_\mathrm{red}$). In the next step, we rescaled all uncertainties to get $\chi^{2}_\mathrm{red} = 1$. $S/N$ of the spectra can be obtained from the $1\sigma$ uncertainties as $C/\sigma$, where the scaling constant $C$ was found to be about 1.0 km~s$^{-1}$. Spectra were also analyzed using the cross-correlation function (CCF) technique, see \citet{Baranne1}. The data were cross-correlated with a K0V stellar mask, with an effective temperature of $T_\mathrm{eff} = 5250$ K, a surface gravity of $\log g = 4.5$ [cgs], and solar metallicity [Fe/H] = 0.0 dex. To get RVs we used only a spectral region around the magnesium triplet, i.e., between the wavelengths of 4900 and 5400 \AA, where no telluric lines, nor hydrogen Balmer lines are present. The advantage of the BF solution is generally smaller uncertainty of individual RVs, while the CCF technique is more consistent with the discovery RVs (see Sect. \ref{refsys}).  An overview of the obtained dataset is shown in Tab. \ref{spectroobslog}, where we present BF RVs, as well as CCF RVs. The first observation was taken at $\mathrm{BJD_{TDB}} = 2~458~928.385194$, which corresponds to 21:13:34.28 UT on March 19, 2020, and the last observation ($\mathrm{BJD_{TDB}} = 2~460~153.455588$) is from 22:57:04.96 UT on July 27, 2023. This means that the timebase of the RVs is about 1225 days, fully covering and exceeding the \textit{TESS} observations window (compare with Tab. \ref{TESSobslog}). Unfortunately, we had to discard a couple of spectroscopic observations from the analysis due to problems with the stability of the spectrograph or due to outlier data points.

For the purpose of spectroscopic analysis of the planet's host star the obtained 1D spectra were combined into one spectrum to increase the signal-to-noise ratio using \texttt{iSpec}\footnote{See \url{https://www.blancocuaresma.com/s/iSpec}.} \citep{Blanco1, Blanco2}. We shifted all of the 20 obtained spectra into the rest frame and combined the barycentric correction and the intrinsic RV correction into one step applying the \texttt{iSpec} cross-correlation routine. We cross-correlated the spectra with a template from the \citet{Munari1} synthetic spectrum library. According to the literature values for the stellar parameters, i.e., $T_\mathrm{eff} = 6250 \pm 100$ K, $\log g = 4.246 \pm 0.023$ [cgs], and [Fe/H] = $0.432 \pm 0.057$ dex \citep{Crouzet1}, we selected a template file from the spectrum library, which corresponds to $T_\mathrm{eff} = 6250$ K, $\log g = 4.5$ [cgs] and [Fe/H] = $+0.5$ dex. After shifting the spectra into the rest frame we averaged them via median into a final spectrum by setting the resolution to $R = 35~000$ and setting the sampling to 0.05 \AA. We then slightly shifted the overall continuum level to set it to 1.0 as much as possible. In this way, we obtained the final averaged spectrum for the planet's host star XO-7, which we further analyzed in order to obtain fundamental stellar parameters (see Sect. \ref{spectrsynthres}).  

\section{The planet's host star}
\label{spectrsynthres}

\begin{table}
\centering
\caption{An overview of fundamental facts about the planet's host star. The table includes stellar parameters obtained in this work from the spectra of XO-7, as well. Notes: $^{\star}$Derived from the redder region of the final spectrum, see the text in Sect. \ref{spectrsynthres}. $^{\star\star}$Derived from the bluer region of the final spectrum, see the text in Sect. \ref{spectrsynthres}. C2020 = \citet{Crouzet1}, A1903 = \citet{Argelander1}, G2023 = \citet{Gaia2}, G2021 = \citet{Gaia1}, B2018 = \citet{Bailer1}, H2000 = \citet{Hog1}, C2003 = \citet{Cutri1}.}
\label{spectrares}
\begin{tabular}{lll}
\hline
Parameter [unit] & Value & Source\\
\hline
Name & XO-7 & C2020\\
Catalog Name & BD+85 317 & A1903\\
\textit{Gaia} DR3 ID & 2303332931542914048 & G2023\\
RA (J2000.0) & 18:29:54.9 & G2021\\
Dec (J2000.0) & +85:13:59 & G2021\\
$p$ [mas] & $4.3216 \pm 0.0132$ & G2023\\
$\mu_\alpha$ [$\mathrm{mas~yr^{-1}}$] & $-15.208 \pm 0.015$ & G2023\\
$\mu_\delta$ [$\mathrm{mas~yr^{-1}}$] & $24.396 \pm 0.017$ & G2023\\ 
$D$ [pc] & $234.1 \pm 1.2$ & B2018\\
$B$ [mag] & $11.23 \pm 0.06$ & H2000\\    
$V$ [mag] & $10.52 \pm 0.04$ & H2000\\
$G$ (\textit{Gaia}) [mag] & $10.459 \pm 0.002$ & G2021\\ 
$J$ [mag] & $9.557 \pm 0.024$ & C2003\\
$H$ [mag] & $9.308 \pm 0.030$ & C2003\\
$K$ [mag] & $9.241 \pm 0.024$ & C2003\\
Spectral type & G0V & C2020\\
$T_\mathrm{eff}$ [K] & $6250 \pm 100$ & C2020\\
$T_\mathrm{eff}$ [K] & $5957 \pm 388^{\star}$/$6092 \pm 164^{\star\star}$ & This work\\
$\log g$ [cgs] & $4.246 \pm 0.023$ & C2020\\
$\log g$ [cgs] & $4.06 \pm 0.67^{\star}$/$3.76 \pm 0.39^{\star\star}$ & This work\\
Fe/H [dex] & $0.432 \pm 0.057$ & C2020\\
Fe/H [dex] & $0.39 \pm 0.22^{\star}$/$0.42 \pm 0.13^{\star\star}$ & This work\\
$v \sin i$ [$\mathrm{km~s}^{-1}$] & $6.00 \pm 1.00$ & C2020\\
$v \sin i$ [$\mathrm{km~s}^{-1}$] & $9.75 \pm 1.45^{\star}$/$12.04 \pm 1.07^{\star\star}$ & This work\\   
$M_\mathrm{s}$ [$M_\odot$] & $1.405 \pm 0.059$ & C2020\\
$R_\mathrm{s}$ [$R_\odot$] & $1.480 \pm 0.022$ & C2020\\
Age [Gyr] & $1.18^{+0.98}_{-0.71}$ & C2020\\
\hline
\end{tabular}
\end{table}

The final averaged spectrum of XO-7 was fitted using the \texttt{iSpec} fitting routine, which performs spectral synthesis. We selected two spectral regions for fitting: as the first (redder) region we selected a wavelength range from 4895 to 5553 \AA. As the second (bluer) region with hydrogen Balmer lines from 4220 to 5042 \AA~ was selected and fitted. During the fitting procedure, we fixed the limb-darkening coefficient $u$ to 0.5, the stellar microturbulent velocity $v_\mathrm{mic}$ to 1.0 $\mathrm{km~s}^{-1}$, and the stellar macroturbulent velocity $v_\mathrm{mac}$ to zero, assuming no motion in larger atmospheric cells. We freely adjusted the following fundamental stellar parameters: $T_\mathrm{eff}$, $\log g$, [Fe/H], and the projected rotational velocity $v \sin i$.

\begin{figure}
\includegraphics[width=\columnwidth]{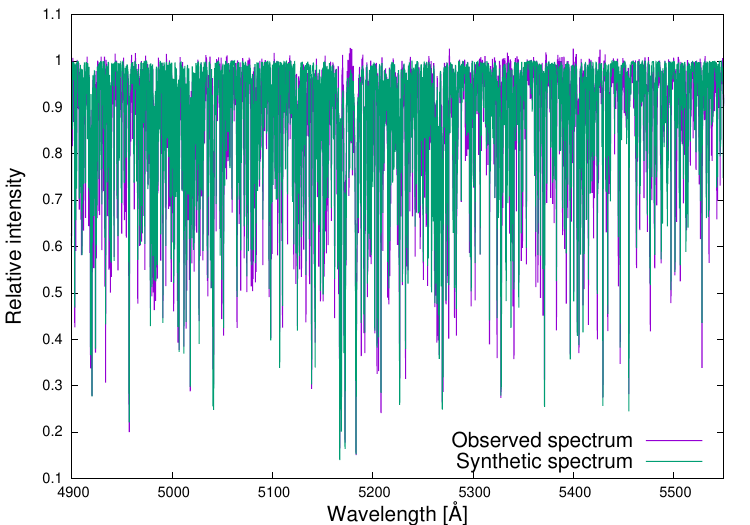}
\caption{The final averaged spectrum of the host star XO-7, overplotted with the synthetic spectrum in the redder region.}
\label{xo7spectrum} 
\end{figure}

The obtained results are included in Tab. \ref{spectrares}. The observed and averaged stellar spectrum, overplotted with the synthetic spectrum in the redder region is depicted, as an illustrative example, in Fig. \ref{xo7spectrum}. Although we did not improve these parameters compared to the discovery values, the final spectrum of XO-7 reveals well the fundamental characteristics of the planet's host star. Based on the data listed in Tab. \ref{spectrares} we can see that the best-fitting values obtained from the bluer region are more precise. The parameter values are in a $3\sigma$ agreement with the corresponding values found by \citet{Crouzet1}, except for the $v \sin i$ parameter value derived from the bluer region of the final spectrum, which is $12.04 \pm 1.07$ $\mathrm{km~s}^{-1}$, while the discoverers presented a value of $v \sin i = 6.0 \pm 1.0$ $\mathrm{km~s}^{-1}$. The difference, in this case, is about $5.6 \sigma$, which we can explain with different spectral resolutions of the spectrographs (see Sections \ref{spectroobs} and \ref{refsys}).

\begin{table*}
\centering
\caption{An overview of the \texttt{Allesfitter} best-fitting parameters of the exoplanet system XO-7 obtained from the \textit{TESS} 2-min PDCSAP photometry and RV observations. Notes: $^1$Taken from, or calculated based on \citet{Crouzet1}. $^2$Based on the stellar parameters of $T_\mathrm{eff} = 6250$ K and $\log g $ = 4.246 [cgs] \citep{Crouzet1}. $^3$Shifted automatically by the \texttt{Allesfitter} software package.}
\label{newparameterstab}
\begin{tabular}{lllll}
\hline
Parameter [unit] & Description & Prior & Value\\
\hline
$T_\mathrm{c}$ [$\mathrm{BJD}_\mathrm{TDB}$] & Reference mid-transit time & $\mathcal{N}$(2457917.47503, 0.00045)$^1$ & $2458779.58040 \pm 0.00016^3$\\
$P_\mathrm{orb}$ [d] & Orbital period & $\mathcal{N}$(2.8641424, 0.0000043)$^1$ & $2.86413296 \pm 0.00000055$\\
$R_\mathrm{p}/R_\mathrm{s}$ & Planet-to-star radius ratio & $\mathcal{N}$(0.09532, 0.00093)$^1$ & $0.09344 \pm 0.00028$\\
$(R_\mathrm{p} + R_\mathrm{s})/a$ & Scaled sum of fractional radii & $\mathcal{N}$(0.1703, 0.0037)$^1$ & $0.1739 \pm 0.0013$\\
$\cos i$ & Cosine of the orbit inclination angle & $\mathcal{N}$(0.1140, 0.0050)$^1$ & $0.1179 \pm 0.0017$\\
$K$ [$\mathrm{km~s^{-1}}$] & RV semi-amplitude & $\mathcal{N}$(0.0805, 0.0032)$^1$ & $0.0805 \pm 0.0021$\\
$q_1$ & Limb darkening coefficient & $\mathcal{U}$(0.2132, 0.4132)$^2$ & $0.291_{-0.026}^{+0.028}$\\
$q_2$ & Limb darkening coefficient & $\mathcal{U}$(0.2098, 0.4098)$^2$ & $0.302_{-0.062}^{+0.069}$\\
$\log \sigma_\mathrm{TESS}$ [$\log$ relative flux] & Instrumental noise, \textit{TESS} data & $\mathcal{U}$(-15.0, 0.0) & $-6.8592 \pm 0.0051$\\ 
$\log \sigma_\mathrm{SOPHIE}$ [$\log \mathrm{km~s^{-1}}$] & Instrumental noise, SOPHIE data & $\mathcal{U}$(-15.0, 0.0) & $-9.2_{-3.8}^{+3.5}$\\
$\log \sigma_\mathrm{MUSICOS}$ [$\log \mathrm{km~s^{-1}}$] & Instrumental noise, MUSICOS data & $\mathcal{U}$(-15.0, 0.0) & $-8.6 \pm 4.3$\\ 
$\log S_\mathrm{0, TESS}$ & Scaled power, \textit{TESS} SHO GP kernel & $\mathcal{U}$(-30.0, 0.0) & $-16.08 \pm 0.30$\\ 
$\log \omega_\mathrm{0, TESS}$ [$\log$ d$^{-1}$] & Frequency, \textit{TESS} SHO GP kernel & $\mathcal{U}$(-15.0, 15.0) & $0.97_{-0.19}^{+0.21}$\\
$O_\mathrm{SOPHIE}$ [$\mathrm{km~s^{-1}}$] & Offset, SOPHIE data & $\mathcal{U}$(-13.15, -12.80) & $-12.9308 \pm 0.0038$\\   
$S_\mathrm{SOPHIE}$ [$\mathrm{km~s^{-1}~T_{obs}^{-1}}$] & Linear slope, SOPHIE data & $\mathcal{U}$(-0.2, 0.0) & $-0.1040 \pm 0.0065$\\
$O_\mathrm{MUSICOS}$ [$\mathrm{km~s^{-1}}$] & Offset, MUSICOS data & $\mathcal{U}$(-14.1, -13.4) & $-13.782 \pm 0.018$\\  
$S_\mathrm{MUSICOS}$ [$\mathrm{km~s^{-1}~T_{obs}^{-1}}$] & Linear slope, MUSICOS data & $\mathcal{U}$(-0.2, 0.2) & $0.040 \pm 0.029$\\
\hline
\end{tabular}
\end{table*}

\begin{table}
\centering
\caption{An overview of the \texttt{Allesfitter} derived parameters of the exoplanet system XO-7 obtained from the \textit{TESS} 2-min PDCSAP photometry and RV observations. Notes: $^{\dagger}$Mass ratio. $^{\star}$Total transit duration between the 1st and the 4th contact. $^{\star\star}$Full transit duration between the 2nd and the 3rd contact. $^{\diamond}$Assuming $T_\mathrm{eff} = 6250 \pm 100~\mathrm{K}$, $R_\mathrm{s} = 1.480 \pm 0.022~\mathrm{R}_\odot$ \citep{Crouzet1}, $a = 9.31 \pm 0.16~\mathrm{R}_\odot$, the albedo of 0.3, and uniform heat redistribution. $^{\ddagger}$Transit depth.}
\label{newparameterstab2}
\begin{tabular}{ll}
\hline
Parameter [unit] & Value\\
\hline
$R_\mathrm{s}/a$ & $0.1590 \pm 0.0012$\\
$a/R_\mathrm{s}$ & $6.289 \pm 0.046$\\ 
$R_\mathrm{p}/a$ & $0.01486 \pm 0.00014$\\ 
$R_\mathrm{p}$ [$\mathrm{R_{\oplus}}$] & $15.08 \pm 0.23$\\ 
$R_\mathrm{p}$ [$\mathrm{R_{Jup}}$] & $1.346 \pm 0.020$\\ 
$a$ [$\mathrm{R_{\odot}}$] & $9.31 \pm 0.16$\\ 
$a$ [au] & $0.04329 \pm 0.00072$\\ 
$i$ [deg] & $83.228 \pm 0.098$\\ 
$q_\mathrm{m}^{\dagger}$ & $0.000493 \pm 0.000015$\\
$M_\mathrm{p}$ [$\mathrm{M_{\oplus}}$] & $231 \pm 12$\\ 
$M_\mathrm{p}$ [$\mathrm{M_{Jup}}$] & $0.726 \pm 0.038$\\ 
$M_\mathrm{p}$ [$\mathrm{M_{\odot}}$] & $0.000693 \pm 0.000036$\\ 
$b$ & $0.7416 \pm 0.0056$\\ 
$T_\mathrm{14}^{\star}$ [h] & $2.8232 \pm 0.0078$\\ 
$T_\mathrm{23}^{\star\star}$ [h] & $1.829 \pm 0.016$\\ 
$\rho_\mathrm{p}$ [g~cm$^{-3}$] & $0.369_{-0.028}^{+0.030}$\\ 
$g_\mathrm{p}$ [m~s$^{-2}$] & $9.32 \pm 0.29$\\ 
$T_\mathrm{eq}^{\diamond}$ [K] & $1612 \pm 27$\\ 
$T_\mathrm{d}^\ddagger$ [relative flux] & $0.008810_{-0.000028}^{+0.000026}$\\ 
$u_\mathrm{1}$ & $0.326_{-0.059}^{+0.063}$\\ 
$u_\mathrm{2}$ & $0.213 \pm 0.079$\\ 
$\rho_\mathrm{s}$ [g~cm$^{-3}$] & $0.574 \pm 0.013$\\ 
\hline
\end{tabular}
\end{table}

\section{Data analysis and results}
\label{dataanres}

\subsection{Refined system parameters}                
\label{refsys}

Orbital and planetary parameters of the XO-7 system were obtained so far by \citet{Crouzet1} from ground-based multicolor photometric data and RV observations, and by \citet{Maciejewski3} from three sectors \textit{TESS} photometric data alone. In order to derive refined orbital and planetary parameters of the system, we combined all 2-min integrated \textit{TESS} PDCSAP photometric data of XO-7 and RV observations. In this case, we used the 20 BF RV measurements described in Sect. \ref{spectroobs} and listed in Tab. \ref {spectroobslog} (these data have smaller uncertainties), and the RV observations published by \citet{Crouzet1}. These RVs represent 49 observations obtained between $\mathrm{BJD_{TDB}} = 2~457~593.3798$, which corresponds to 21:08:02.39 UT on July 23, 2016, and $\mathrm{BJD_{TDB}} = 2~458~303.5649$, which is 01:34:55.27 UT on July 04, 2018, with the SOPHIE spectrograph \citep{Perruchot1} installed on the 1.93 m telescope of Observatoire de Haute-Provence (OHP) in France. The timebase of these observations is about 710 days. The authors used its High-Resolution (HR) mode, which gives a resolving power of $R = 75~000$. The exposure times were around 13 min allowing $S/N$ of around 27 per pixel at 5500 \AA. The RV data were determined using the CCF technique. The authors opted for a K0V stellar mask to cross-correlate the data. The list of the RV observations with $\pm 1\sigma$ uncertainties can be found in the appendix of \citet{Crouzet1}, in Tab. 5.              

\begin{figure*}
\centering
\centerline{
\includegraphics[width=\columnwidth]{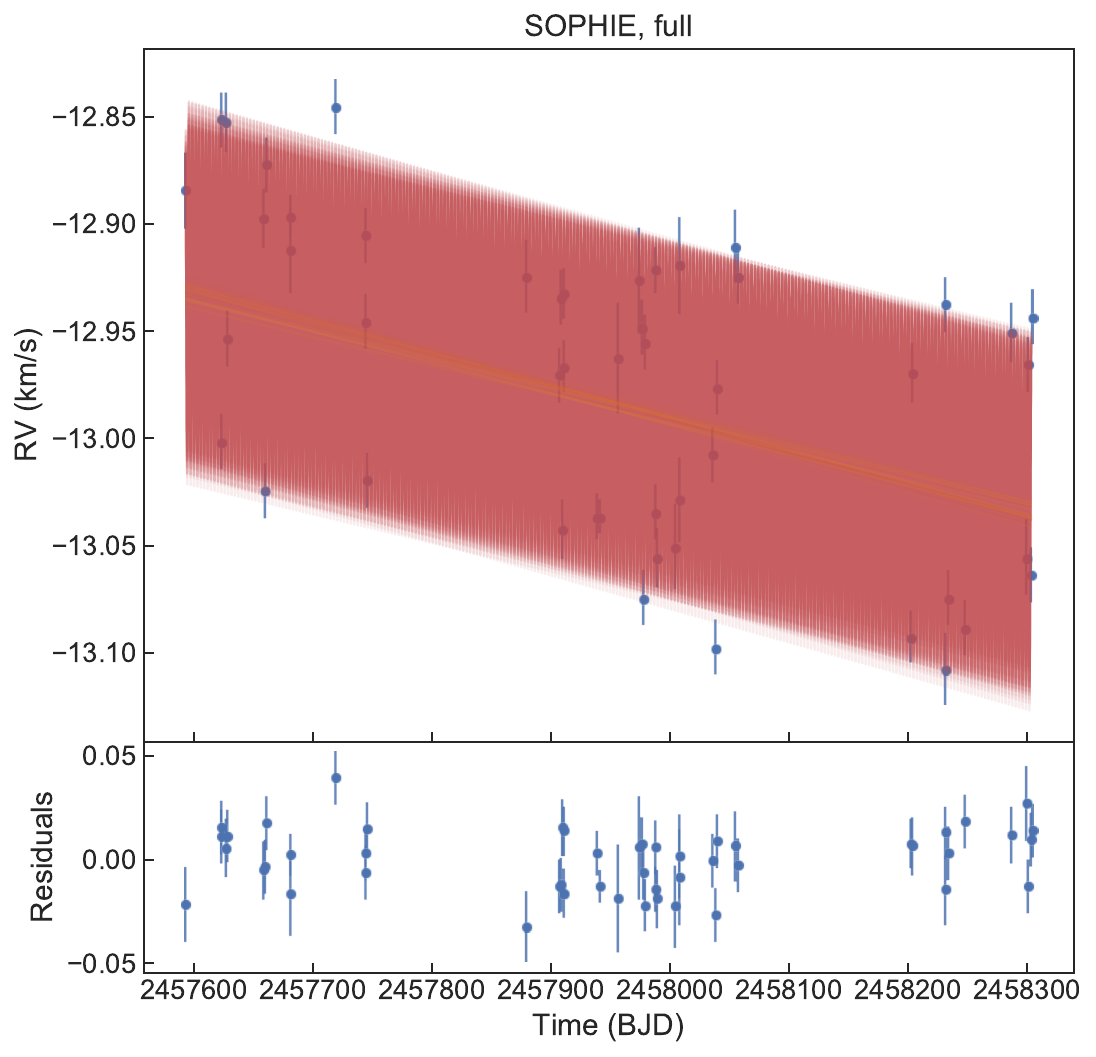}
\includegraphics[width=\columnwidth]{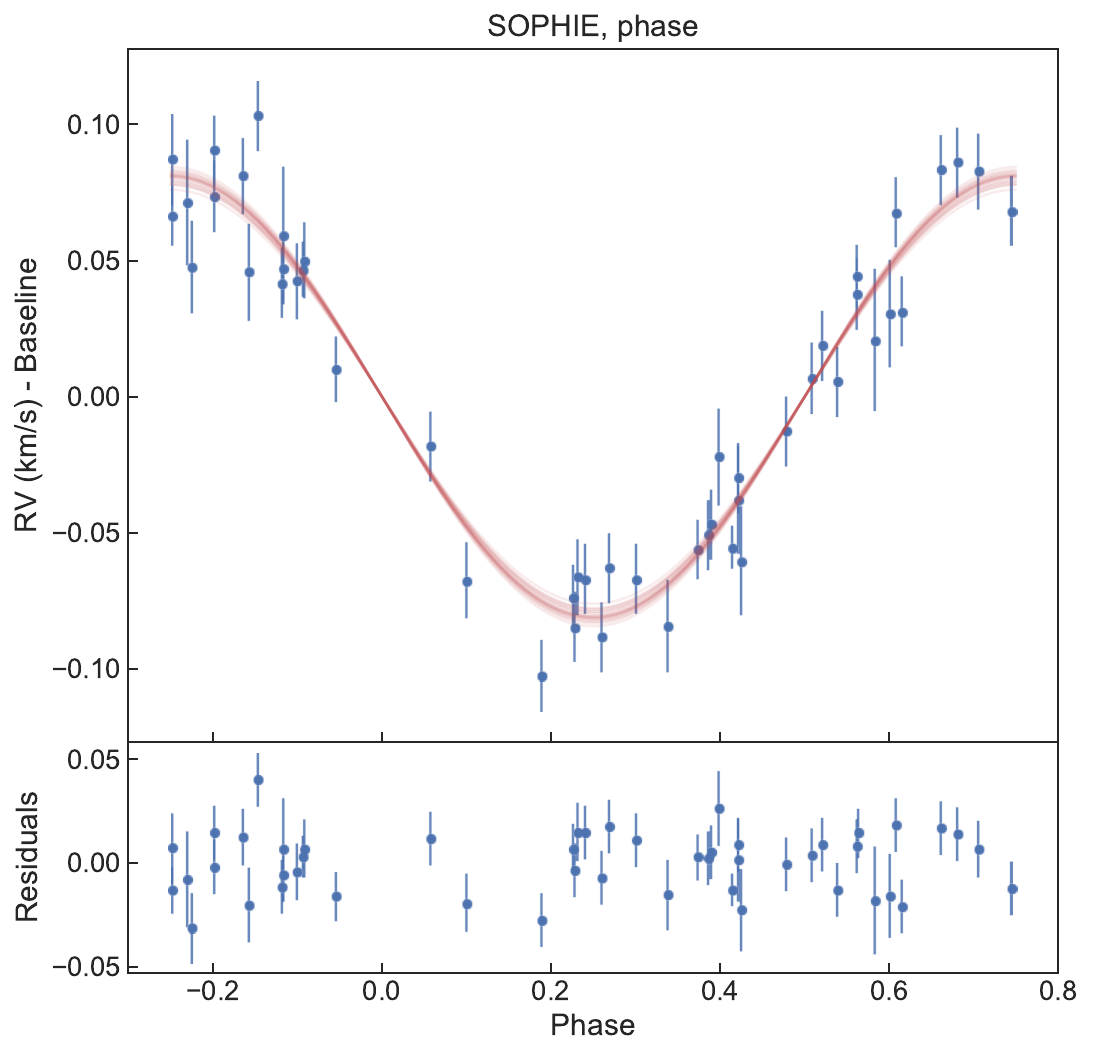}}
\centerline{
\includegraphics[width=89mm]{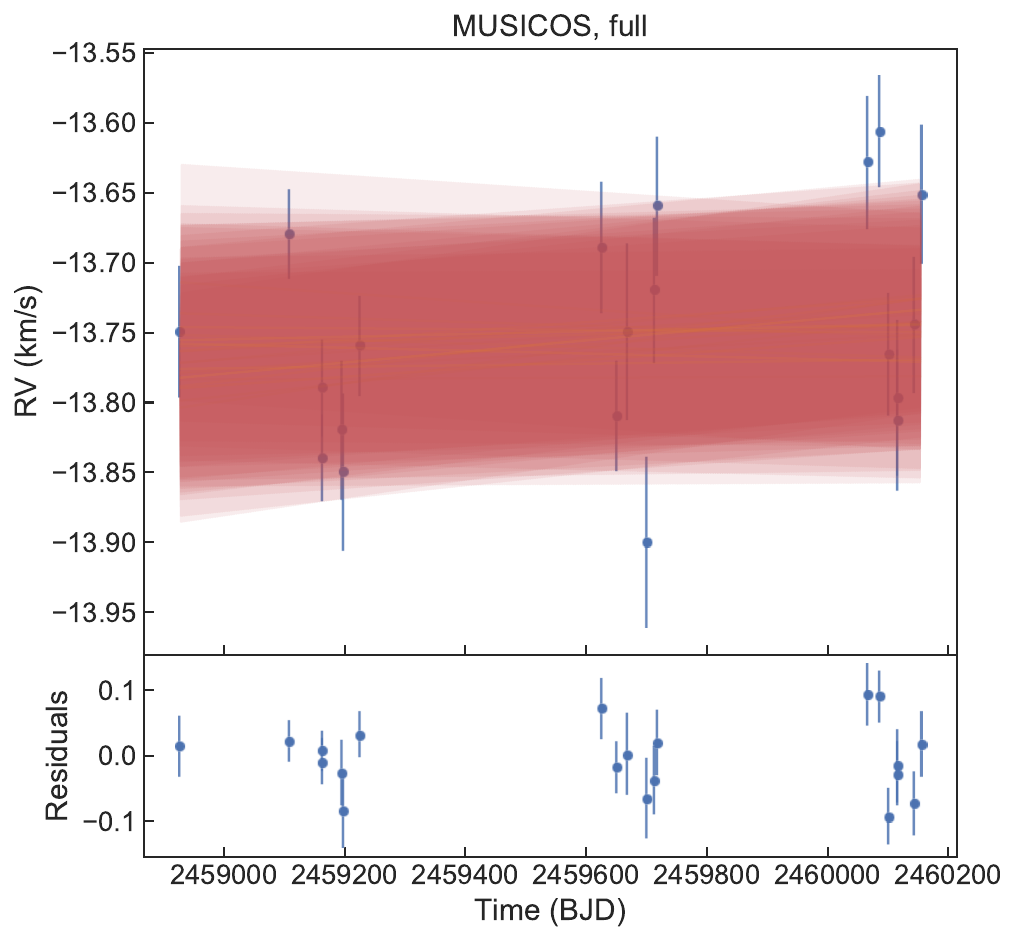}
\includegraphics[width=\columnwidth]{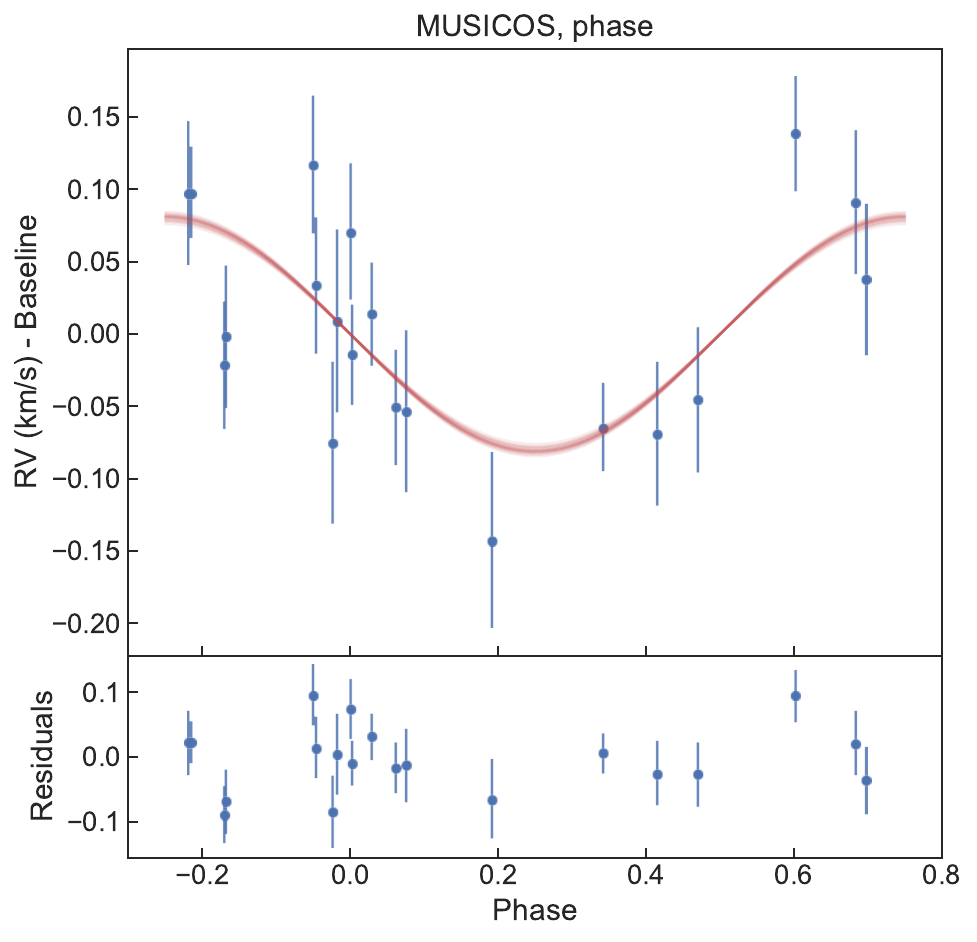}}
\caption{SOPHIE (top panels) and MUSICOS (bottom panels) RV observations of XO-7, overplotted with the best-fitting \texttt{Allesfitter} model (20 curves from random posterior samples). In the case of MUSICOS, we used BF RVs. Without phase-folding (left-hand panels) to see the linear slope in the RV data and with phase-folding (right-hand panels) to see the RV amplitude. Residuals are also shown. During this modeling procedure, all available 2-min integrated \textit{TESS} PDCSAP data and the RV observations were fitted simultaneously.}
\label{RVplotSOPHIE+MUSICOS} 
\end{figure*} 

\begin{figure}
\includegraphics[width=\columnwidth]{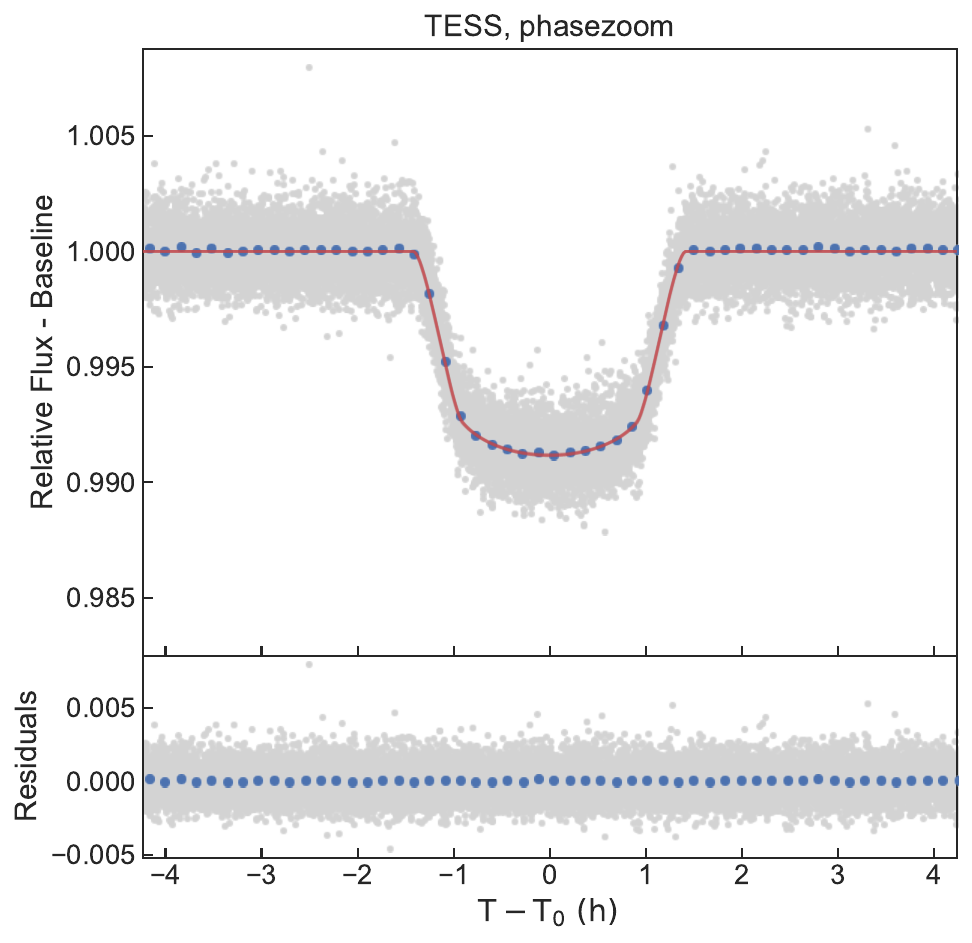}
\caption{Stacked and binned \textit{TESS} transit light curve of XO-7b, overplotted with the best-fitting \texttt{Allesfitter} model (20 curves from random posterior samples). Residuals are also shown. During this modeling procedure, all available 2-min integrated \textit{TESS} PDCSAP data and the RV observations were fitted simultaneously.}
\label{TESSplotREFINED} 
\end{figure} 

To derive new orbital and planetary parameters we employed the \texttt{Allesfitter}\footnote{See \url{https://www.allesfitter.com}.} software package \citep{Guenther1, Guenther2}. It is a public software for modeling photometric and RV data. It can accommodate multiple exoplanets, multi-star systems, starspots, stellar flares, TTVs, and various noise models. \texttt{Allesfitter} automatically runs a nested sampling or Markov Chain Monte Carlo (MCMC) fit. For all this, it constructs an inference framework that unites the versatile packages \texttt{ellc} \citep{Maxted1}, \texttt{aflare} \citep{Davenport1}, \texttt{dynesty} \citep{Speagle1}, \texttt{emcee} \citep{Foreman1} and \texttt{celerite} \citep{Kallinger1, Foreman2, Barros1}. In order to construct the combined \textit{TESS} and RV model we opted for the nested sampling fit option with initial settings. To speed up the computation process we cut the \textit{TESS} light curves and selected only the regions with a data window width of 0.4 d, centered on the mid-transit points. Several parameters were optimized during the fitting procedure, including the reference mid-transit time $T_\mathrm{c}$, the orbital period $P_\mathrm{orb}$, the planet-to-star radius ratio $R_\mathrm{p}/R_\mathrm{s}$, the scaled sum of fractional radii $(R_\mathrm{p} + R_\mathrm{s})/a$, the cosine of the orbit inclination angle ($\cos i$), and the RV semi-amplitude $K$. The quadratic limb darkening (LD) law was applied during the fitting procedure. The $u_1$ and $u_2$ LD coefficients were first linearly interpolated based on the stellar parameters of $T_\mathrm{eff} = 6250$ K and $\log g $ = 4.246 [cgs] \citep{Crouzet1} from the tables of coefficients calculated for the \textit{TESS} passband using the \texttt{PHOENIX-COND} models by \citet{Claret1}. We then converted these LD coefficients to $q_1$ and $q_2$ \citep{Kipping1} and during the optimization procedure, we applied $\pm 0.1$-wide uniform priors centered on these values to avoid the non-physical solution of the light curve. Since trends due to stellar and instrumental noise were still present in the PDCSAP photometry, in order to model the \textit{TESS} flux baseline we applied a Gaussian Process (GP) regression method using the \texttt{SHOTerm} (Simple Harmonic Oscillator -- SHO) plus \texttt{JitterTerm} kernel, implemented in the \texttt{celerite}\footnote{See \url{https://celerite.readthedocs.io/en/stable/}.} package, with a fixed quality factor of $Q_0 = 1/\sqrt{2}$, as it is common for quasi-periodic stellar variability. The regression is done by using $\log \sigma$ (free), $\log Q_0$ (fixed), $\log \omega_0$ (free), and $\log S_0$ (free) hyperparameters with bounds on the values of these parameters to be inputted by the user. The RV baseline was modeled with a linear function motivated by the discoverers, who reported on a linear slope in the discovery RV data with a value of $S_\mathrm{SOPHIE} = -0.1480 \pm 0.0010~\mathrm{m~s^{-1}~d^{-1}}$. The instrumental noise in the RV data was sampled with the \texttt{JitterTerm} kernel, using the $\log \sigma$ (free) hyperparameter. We assumed a circular orbit of XO-7b. The list of fitted parameters is presented in Tab. \ref{newparameterstab} and the derived parameters are listed in Tab. \ref{newparameterstab2}. The fitted RVs are depicted in Fig. \ref{RVplotSOPHIE+MUSICOS} and the stacked and binned \textit{TESS} PDCSAP transit light curve of XO-7b, overplotted with the best-fitting \texttt{Allesfitter} model is shown in Fig. \ref{TESSplotREFINED}. We can clearly see that the previously observed significant linear slope in the SOPHIE RVs was not confirmed with the follow-up RV data. The linear slope in the case of the MUSICOS BF RV observations is only marginal. We discuss this interesting result in Sect. \ref{discuss}. 

XO-7b is a hot Jupiter on a close-in orbit with a semi-major axis of $a = 0.04329 \pm 0.00072$ au and with a short orbital period below 10 days. We see its orbit nearly edge-on with an inclination angle of $i = 83.228 \pm 0.098$ deg, which corresponds to an impact parameter of $b = 0.7416 \pm 0.0056$. The total transit duration between the 1st and the 4th contact is $T_\mathrm{14} = 2.8232 \pm 0.0078$ h. Using all available 2-min integrated \textit{TESS} PDCSAP data we significantly improved the linear ephemeris of the planet compared to the previous values presented by \citet{Crouzet1} and \citet{Maciejewski3}. The new linear ephemeris, obtained from the \texttt{Allesfitter} combined model, is:

\begin{equation}
\begin{split}
\label{linef1}
T_0 = T_\mathrm{c} + P_\mathrm{orb} \times E =\\
&\hspace{-28mm} = 2~458~779.58040 \pm 0.00016~\mathrm{BJD_{TDB}} +\\
&\hspace{-28mm} + 2.86413296 \pm 0.00000055~\mathrm{d} \times E,
\end{split}
\end{equation}

\noindent where $T_0$ is the mid-transit time of an arbitrary transit and $E$ is the epoch of observation, i.e., the number of the orbital cycle calculated from the reference mid-transit time $T_\mathrm{c}$. XO-7b is an inflated planet because it has a radius larger than the radius of Jupiter, but its mass is less than the mass of Jupiter. We obtained a planet-to-star radius ratio, which is $R_\mathrm{p}/R_\mathrm{s} = 0.09344 \pm 0.00028$. Using a stellar radius of $R_\mathrm{s} = 1.480 \pm 0.022~\mathrm{R}_\odot$ \citep{Crouzet1}, we can get an absolute planet radius of $R_\mathrm{p} = 1.346 \pm 0.020~\mathrm{R_{Jup}}$. The mass of the planet was derived based on an RV semi-amplitude, which is $K = 0.0805 \pm 0.0021~\mathrm{km~s^{-1}}$. This gives $M_\mathrm{p} = 0.726 \pm 0.038~\mathrm{M_{Jup}}$. Using a stellar mass of $M_\mathrm{s} = 1.405 \pm 0.059~\mathrm{M}_\odot$ \citep{Crouzet1} we can get a mass ratio of $q_\mathrm{m} = 0.000493 \pm 0.000015$ and applying the obtained planet mass and radius parameter values, we can get a planet density of $\rho_\mathrm{p} = 0.369_{-0.028}^{+0.030}$ g~cm$^{-3}$. This is only about 28.5\% of Jupiter's density. Based on its mass and radius a very similar planet is HAT-P-30b/WASP-51b \citep{Johnson1, Enoch1}.  

\subsection{TTV analysis}
\label{ttvanalysis}

\begin{table}
\scriptsize
\centering
\caption{The list of the O-C values of mid-transit times of XO-7b derived from the 2-min integrated \textit{TESS} PDCSAP data using the \texttt{Allesfitter} combined model parameter values (see Tab. \ref{newparameterstab}). The epoch $E$ was counted from $T_\mathrm{c} = 2~458~779.58040~\mathrm{BJD_{TDB}}$.}
\label{ttvfitdata}
\begin{tabular}{cccc}
\hline
Transit & $E$ & O-C [d] & $\pm 1\sigma$ [d]\\										
\hline
No. 1  &  72 & $-0.00016$ & 0.00062\\             
No. 2  &  73 & $+0.00042$ & 0.00058\\             
No. 3  &  74 & $-0.00003$ & 0.00063\\             
No. 4  &  75 & $+0.00027$ & 0.00062\\             
No. 5  &  76 & $+0.00007$ & 0.00058\\             
No. 6  &  77 & $+0.00026$ & 0.00064\\             
No. 7  &  78 & $-0.00046$ & 0.00059\\             
No. 8  &  79 & $+0.00023$ & 0.00061\\             
No. 9  &  80 & $+0.00044$ & 0.00060\\             
No. 10 &  81 & $-0.00129$ & 0.00060\\             
No. 11 &  82 & $-0.00017$ & 0.00064\\             
No. 12 &  83 & $-0.00044$ & 0.00060\\             
No. 13 &  84 & $+0.00047$ & 0.00061\\             
No. 14 &  86 & $-0.00094$ & 0.00063\\             
No. 15 &  87 & $-0.00039$ & 0.00062\\             
No. 16 &  88 & $-0.00110$ & 0.00067\\             
No. 17 &  89 & $-0.00010$ & 0.00059\\             
No. 18 & 214 & $+0.00004$ & 0.00049\\             
No. 19 & 215 & $+0.00022$ & 0.00049\\             
No. 20 & 216 & $+0.00156$ & 0.00048\\             
No. 21 & 217 & $-0.00077$ & 0.00048\\             
No. 22 & 218 & $+0.00029$ & 0.00048\\             
No. 23 & 219 & $+0.00060$ & 0.00050\\             
No. 24 & 220 & $-0.00080$ & 0.00050\\             
No. 25 & 221 & $+0.00112$ & 0.00048\\             
No. 26 & 222 & $+0.00052$ & 0.00050\\             
No. 27 & 223 & $-0.00051$ & 0.00047\\             
No. 28 & 280 & $-0.00018$ & 0.00049\\             
No. 29 & 281 & $-0.00043$ & 0.00048\\             
No. 30 & 282 & $-0.00021$ & 0.00047\\             
No. 31 & 283 & $-0.00070$ & 0.00048\\             
No. 32 & 285 & $-0.00034$ & 0.00049\\             
No. 33 & 286 & $+0.00015$ & 0.00046\\             
No. 34 & 287 & $-0.00029$ & 0.00049\\             
No. 35 & 288 & $+0.00050$ & 0.00046\\             
No. 36 & 328 & $-0.00024$ & 0.00049\\             
No. 37 & 329 & $+0.00027$ & 0.00050\\             
No. 38 & 330 & $+0.00035$ & 0.00051\\             
No. 39 & 331 & $-0.00024$ & 0.00050\\             
No. 40 & 333 & $-0.00011$ & 0.00049\\             
No. 41 & 334 & $-0.00018$ & 0.00053\\             
No. 42 & 335 & $-0.00074$ & 0.00051\\             
No. 43 & 336 & $-0.00048$ & 0.00048\\             
No. 44 & 337 & $+0.00024$ & 0.00053\\             
No. 45 & 338 & $+0.00006$ & 0.00053\\             
No. 46 & 339 & $-0.00131$ & 0.00058\\             
No. 47 & 340 & $+0.00040$ & 0.00048\\             
No. 48 & 342 & $+0.00082$ & 0.00055\\             
No. 49 & 343 & $-0.00004$ & 0.00050\\
No. 50 & 344 & $+0.00071$ & 0.00054\\             
No. 51 & 345 & $+0.00005$ & 0.00049\\             
No. 52 & 396 & $-0.00124$ & 0.00048\\             
No. 53 & 397 & $+0.00101$ & 0.00049\\             
No. 54 & 398 & $-0.00059$ & 0.00045\\             
No. 55 & 399 & $+0.00014$ & 0.00047\\             
No. 56 & 401 & $-0.00066$ & 0.00049\\             
No. 57 & 402 & $-0.00026$ & 0.00049\\             
No. 58 & 403 & $-0.00091$ & 0.00048\\             
No. 59 & 405 & $+0.00017$ & 0.00052\\             
No. 60 & 406 & $-0.00031$ & 0.00050\\             
No. 61 & 407 & $-0.00049$ & 0.00051\\             
No. 62 & 408 & $-0.00032$ & 0.00049\\             
No. 63 & 411 & $+0.00101$ & 0.00046\\             
No. 64 & 412 & $+0.00031$ & 0.00050\\             
No. 65 & 413 & $+0.00020$ & 0.00048\\             
\hline
\end{tabular}
\end{table}

\begin{figure}
\includegraphics[width=\columnwidth]{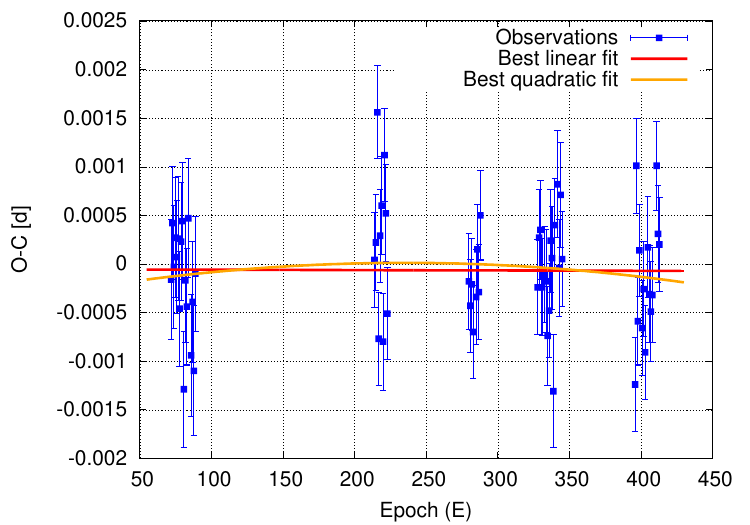}
\caption{Observed-minus-calculated (O-C) diagram of XO-7b mid-transit times, obtained based on 2-min integrated \textit{TESS} PDCSAP data, overplotted with the best-fitting \texttt{OCFIT} models. For more details see Sect. \ref{ttvanalysis}.}
\label{ttvfig} 
\end{figure} 

The first TTV analysis of XO-7b was performed by the discoverers. \citet{Crouzet1} found that TTVs of XO-7b should be lower than 5 min at $1\sigma$ and 15 min at $3\sigma$ over the two years of discovery observations. Based on these photometric measurements \citet{Crouzet1} could rule out the presence of companions massive and close enough to induce significant TTVs on shorter timescales. Later, \citet{Maciejewski3} analyzed \textit{TESS} data from three sectors, Nos. 25, 26, and 40, and found no additional planet in the XO-7 system down to sub-Neptune-sized planets. Since this analysis new \textit{TESS} data were obtained, we decided to repeat the TTV analysis with a much longer timebase, which enabled us to focus on long-term dynamical signs of the announced companion of XO-7b. This could be a sinusoidal modulation of the TTV data, or rather a fraction of this sinusoid, which could be detectable as a quadratic trend in the TTV data. For this purpose, we used the \textit{TESS} data described in Sect. \ref{tessphotdata}. To perform the analysis we again employed the \texttt{Allesfitter} software package described in Sect. \ref{refsys}. We used its TTV function to fit the individual \textit{TESS} transits of XO-7b. In this case, the transit ephemeris is fixed. In addition, we keep fixed the transit shape and the noise model. It means that during this procedure we fixed every parameter to its best value from the \texttt{Allesfitter} combined model (see Tab. \ref{newparameterstab}). Only the observed-minus-calculated (O-C) parameters for individual mid-transit times were fitted. We applied uniform priors on the O-C values, $\mathcal{U}$(-0.025, 0.025) d, and again opted for the nested sampling fit option with initial settings. To speed up the computation process we cut the \textit{TESS} light curves and selected only the regions with a data window width of 0.4 d, centered on the mid-transit points. We excluded partial transits from the dataset. It means that we used 65 \textit{TESS} transits of XO-7b in total, covering a timebase of about 977 days. The calculated mid-transit time of the transit No. 1 in our dataset is $\mathrm{BJD_{TDB}} = 2~458~985.79797$, and the last transit No. 65 has the calculated mid-transit time of $\mathrm{BJD_{TDB}} = 2~459~962.46731$. The list of the fitted O-C values of mid-transit times obtained from this modeling procedure is presented in Tab. \ref{ttvfitdata}.              

\begin{figure*}
\centering
\centerline{
\includegraphics[width=\columnwidth]{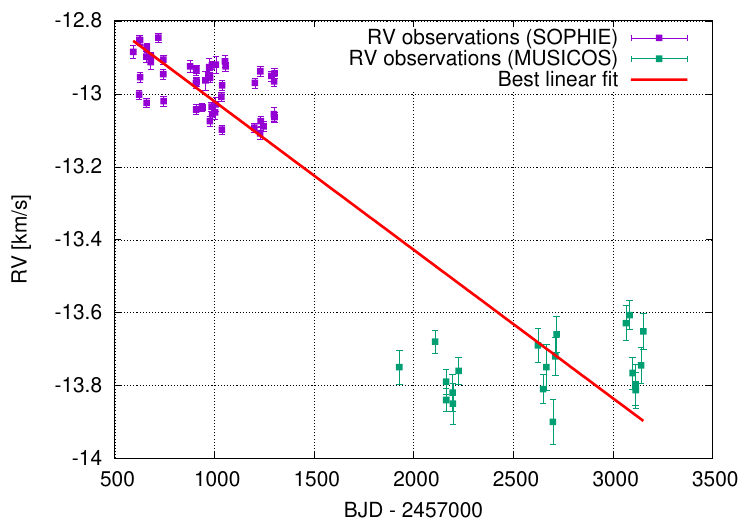}
\includegraphics[width=\columnwidth]{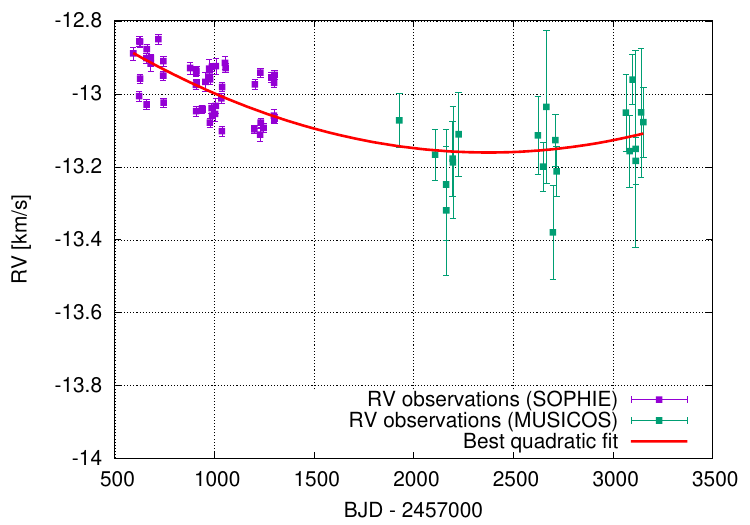}}
\caption{SOPHIE RV observations and MUSICOS BF RVs of XO-7, fitted jointly with a linear model (left-hand panel). SOPHIE RV observations and MUSICOS CCF RVs of XO-7, fitted jointly with a quadratic model (right-hand panel). For the corresponding discussions see the text in Sect. \ref{discuss}.}
\label{RVfit} 
\end{figure*}

The O-C (TTV) data were subsequently inputted in the \texttt{OCFIT}\footnote{See \url{https://github.com/pavolgaj/OCFit}.} code, version 0.2.1 \citep{Gajdos1}, and tested for any quadratic trends in the dataset, which could be an indicator of the announced massive wide-orbit companion. We first fitted the O-C data with a linear function using the \texttt{OCFIT} package \texttt{FitLinear}. The free parameters of the linear model are the reference mid-transit time $T_\mathrm{c}$ and the orbital period $P_\mathrm{orb}$. Subsequently, the O-C data were fitted with a quadratic function, within the \texttt{OCFIT} package \texttt{FitQuad}. The free parameters of the quadratic model are the reference mid-transit time $T_\mathrm{c}$, the orbital period $P_\mathrm{orb}$, and the quadratic coefficient $Q$, which follows from the quadratic ephemeris formula of:

\begin{equation}
\label{quadefemeris1}
T_0 = T_\mathrm{c} + P_\mathrm{orb} \times E + Q \times E^2,
\end{equation}    

\noindent where the quadratic coefficient can be expressed as:

\begin{equation}
\label{quadefemeris2}
Q = \frac{1}{2} P_\mathrm{orb} \times \dot{P},
\end{equation}    

\noindent where $\dot{P}$ means the orbital period change with time $t$, i.e. this is the so-called orbital period change rate: $\dot{P} = \mathrm{d}P/\mathrm{d}t$. $\dot{P}$ is a dimensionless quantity, but it can be expressed, e.g., in d~yr$^{-1}$. During the \texttt{OCFIT} fitting procedure we opted for the robust regression method, performing 100~000 iteration steps. The uncertainties in the fitted parameters of $P_\mathrm{orb}$, $T_\mathrm{c}$, and $Q$ were derived within the \texttt{OCFIT} packages \texttt{FitLinear} and \texttt{FitQuad}, applying the covariance matrix method. We measured the quality of the fit using the Bayesian Information Criterion ($BIC$), calculated automatically by the software. The O-C values of XO-7b mid-transit times, overplotted with the best-fitting \texttt{OCFIT} models are depicted in Fig. \ref{ttvfig}.   

We can see that the O-C data of mid-transit times do not show any periodic features. To check the possible long-term trends, i.e., a fraction of a sinusoid in the O-C data, we evaluated the \texttt{OCFIT} models as follows. Based on the linear fit, we obtained a linear ephemeris of: 

\begin{equation}
\begin{split}
\label{linef2}
T_0 = T_\mathrm{c} + P_\mathrm{orb} \times E =\\
&\hspace{-28mm} = 2~458~779.58034 \pm 0.00022~\mathrm{BJD_{TDB}} +\\
&\hspace{-28mm} + 2.86413292 \pm 0.00000075~\mathrm{d} \times E,
\end{split}
\end{equation}

\noindent and a $BIC$ value of 83.6. The quadratic fit resulted in a quadratic ephemeris of:

\begin{equation}
\begin{split}
\label{quadefemeris3}
T_0 = T_\mathrm{c} + P_\mathrm{orb} \times E + Q \times E^2 =\\
&\hspace{-41mm} = 2~458~779.58011 \pm 0.00040~\mathrm{BJD_{TDB}} +\\
&\hspace{-41mm} + 2.8641354 \pm 0.0000037~\mathrm{d} \times E + Q \times E^2,
\end{split}
\end{equation}

\noindent where the quadratic coefficient is $Q = (-5.29 \pm 7.68) \times 10^{-9}$ d. We obtained a dimensionless value of $\dot{P} = (-3.69 \pm 5.36) \times 10^{-9}$, which is equal to $\dot{P} = (-1.34 \pm 1.96) \times 10^{-6}~\mathrm{d~yr^{-1}}$. Since this result is not significant, the quadratic ephemeris, expressed in Eq. \ref{quadefemeris3}, is not justified. This conclusion is also confirmed by a $BIC$ value of 95.04, which is about 13\% larger than in the case of the linear fit. In summary, we did not find any convincing long-term dynamical sign of the announced wide-orbit massive companion of XO-7b in the O-C data. We did not improve the linear ephemeris by this procedure in Eq. \ref{linef2}, therefore we can adopt the linear ephemeris from the \texttt{Allesfitter} combined model, expressed in Eq. \ref{linef1}, as the final ephemeris solution.    

\section{Discussion}
\label{discuss}

The fitted and derived \texttt{Allesfitter} combined model parameter values, presented in Tabs. \ref{newparameterstab} and \ref{newparameterstab2}, respectively, are in general in a $3\sigma$ agreement to those of presented by the discoverers. Several parameter values were also improved in comparison with these authors. However, very interesting is the result, which we obtained in the case of the RV slope, announced by \citet{Crouzet1}. In agreement with the discovery paper, in the case of the SOPHIE RV data, we obtained a very significant ($\sim 16\sigma$) linear slope, which is quantified by the parameter of $S_\mathrm{SOPHIE} = -0.1040 \pm 0.0065~\mathrm{km~s^{-1}~T_{obs}^{-1}}$, which means that the change in the RVs is expressed during the timebase of observations. We can easily convert this parameter to the value of $S_\mathrm{SOPHIE} = -0.1464 \pm 0.0092~\mathrm{m~s^{-1}~d^{-1}}$ if we consider that the time difference between the first and the last SOPHIE RV observation is about 710 days. This parameter value is in perfect agreement with the value presented by \citet{Crouzet1}, which is $S_\mathrm{SOPHIE} = -0.1480 \pm 0.0010~\mathrm{m~s^{-1}~d^{-1}}$. Following this discovery value and calculating with a MUSICOS RVs timebase of 1225 days, we expected a linear slope of $S_\mathrm{MUSICOS} \approx -0.1813~\mathrm{km~s^{-1}~T_{obs}^{-1}}$. However, in the case of the MUSICOS BF RVs, we could not confirm this expectation. In this case, we obtained a linear slope, which is expressed with a value of $S_\mathrm{MUSICOS} = 0.040 \pm 0.029~\mathrm{km~s^{-1}~T_{obs}^{-1}}$, which is equal to $S_\mathrm{MUSICOS} = 0.032 \pm 0.024~\mathrm{m~s^{-1}~d^{-1}}$. This value is far from the expected, therefore we can say that the previously observed significant linear slope in the SOPHIE RVs was not confirmed with the follow-up RV data. We detected only a marginal ($\sim 1.3\sigma$) linear slope in the MUSICOS BF RV data. Moreover, this slope has the opposite trend compared to the SOPHIE RV slope. Based on this result the following explanations are plausible.

The first possibility is that there is no linear slope in the RVs. The significant linear slope, observed in the SOPHIE RVs, was caused by a systematic error. This produced an artificial slope in the SOPHIE RV observations. Although technically it is possible, we do not strongly believe that this is the true reason for the marginal detection in the MUSICOS BF RV data. Secondly, we also should take into consideration that the uncertainties of the MUSICOS BF RV data are in general larger compared to the SOPHIE RV observations. On the other hand, the linear RV slope should be increased linearly with time, so this disadvantage is compensated with a longer timebase of MUSICOS RV data. The expected linear slope based on the SOPHIE RV data was $S_\mathrm{MUSICOS} \approx -0.1813~\mathrm{km~s^{-1}~T_{obs}^{-1}}$, as we presented above. This slope should be detectable with MUSICOS BF RV data, despite the larger uncertainties. Finally, we should also consider the fact that there is a velocity offset between the SOPHIE and MUSICOS BF RV observations, which we can express using the absolute difference in the offset parameter values, i.e., $|O_\mathrm{SOPHIE} - O_\mathrm{MUSICOS}| \approx |12.9 - 13.7| \approx 0.8~\mathrm{km~s^{-1}}$. At first glance, this velocity offset could indicate, for example, that the linear slope is still continuing downward, as it was presented by the discoverers (see Fig. \ref{RVfit}, left-hand panel). However, the velocity offset is only artificial due to the different data reduction procedures. The SOPHIE RVs were extracted using a K0V stellar mask via the CCF method. The MUSICOS BF RVs were obtained using  an F7V-type template star. The offset can result from an incorrect RV of the template star, HD\,222368. In fact, the Simbad database\footnote{See \url{https://simbad.u-strasbg.fr/simbad/}.} gives its RV from 5.00 to 5.98~km~s$^{-1}$ from measurements since 2002 (we used a value of 5.4~km~s$^{-1}$). Therefore, in this case, we can tentatively use the MUSICOS CCF RV data (see Sect. \ref{spectroobs} and Tab. \ref{spectroobslog}), which were extracted using a K0V stellar mask (with laboratory wavelength), as well (these data have larger uncertainties). This situation is depicted in Fig. \ref{RVfit} (right-hand panel). In this case, there is a possibility that both RV datasets were collected near the quadrature position of the announced wide-orbit massive companion, where the RVs are nearly stable. This explanation seems to be the most convincing for us. It is also in agreement with the marginal linear slope detected in the MUSICOS BF RVs.  

The quadratic model depicted in Fig. \ref{RVfit} (right-hand panel) may be used to infer and discuss some preliminary physical parameters of the announced companion of XO-7b. To simplify the situation, we will assume that the companion has a circular orbit. The quadratic model $RV_\mathrm{quad}(t)$ in time $t$ can be described based on \citet{Kipping2} as:

\begin{equation}
\begin{split}
\label{Kippingeq}
RV_\mathrm{quad}(t) = O_\mathrm{SOPHIE} + d(t-\bar{t}) + \frac{1}{2}q(t-\bar{t})^2,\\
&\hspace{-59mm} RV_\mathrm{quad}(t) = O_\mathrm{MUSICOS} + d(t-\bar{t}) + \frac{1}{2}q(t-\bar{t})^2,
\end{split}
\end{equation}

\noindent where $O_\mathrm{SOPHIE}$ is the offset of the SOPHIE RV data, $O_\mathrm{MUSICOS}$ is the offset of the MUSICOS CCF RV data, $d$ is the linear acceleration coefficient, $q$ is the quadratic acceleration coefficient (these coefficients are shared in the model between the two datasets), and $\bar{t}$ is the mean epoch of the merged timebase of SOPHIE and MUSICOS RV data, which is $\bar{t} = 2~458~441.372~\mathrm{BJD_{TDB}}$. The fitted parameters of the model, as well as the corresponding asymptotic standard errors, are presented in Tab. \ref{quadmodeltab}. Furthermore, we calculated the minimum RV semi-amplitude $K_\mathrm{min,3}$ induced by the companion, as the absolute difference between the minimum and maximum RV values of the quadratic-fit curve during the merged timebase of RV observations divided by 2, and we obtained a value of $K_\mathrm{min,3} = (|RV_\mathrm{min} - RV_\mathrm{max}|)/2 = (|13.1610 - 12.8880|)/2 = 0.2730/2 = 0.1365~\mathrm{km~s^{-1}}$. The precision of this measurement is $\pm 0.0005~\mathrm{km~s^{-1}}$. In the next step, the minimum orbital period $P_\mathrm{orb,min,3}$ of the announced companion of XO-7b was calculated following \citet{Kipping2}, and \citet{Pinamonti1} as a function of the quadratic acceleration coefficient $q$:

\begin{equation}
\label{Pinamontieq}
P_\mathrm{orb,min,3} \gtrsim 2\pi \sqrt{\frac{2K_\mathrm{min,3}}{q}},
\end{equation}

\noindent and we obtained $7900 \pm 1660$ days, where the uncertainty follows mainly from the uncertainty of $q$. This result seems to be realistic if we take a look at Fig. \ref{RVfit} (right-hand panel), and if we consider that the time difference between the first SOPHIE RV observation and the last MUSICOS RV observation is about 2560 days. Last, but not least, from $P_\mathrm{orb,min,3}$ and $K_\mathrm{min,3}$, the corresponding 'minimum' minimum mass $(M_3 \sin i)_\mathrm{min}$ of the companion can be derived, e.g., by modifying the formula presented by \citet{Torres1}:

\begin{equation}
\label{Torreseq}
(M_3 \sin i)_\mathrm{min} = 4.919 \times 10^{-3}~K_\mathrm{min,3}~(P_\mathrm{orb,min,3})^{1/3}~(M_\mathrm{s})^{2/3},
\end{equation}

\noindent where $K_\mathrm{min,3}$ is in $\mathrm{m~s^{-1}}$, $P_\mathrm{orb,min,3}$ is in days, $M_\mathrm{s}$ is in $\mathrm{M_\odot}$, and the result is expressed in $\mathrm{M_{Jup}}$. In Eq. \ref{Torreseq} we assumed that $M_3 \ll M_\mathrm{s}$ and that the announced companion has a circular orbit. Using the appropriate parameter values from Tabs. \ref{spectrares} and \ref{quadmodeltab}, we obtained a 'minimum' minimum mass of $(M_3 \sin i)_\mathrm{min} = 16.7 \pm 3.5~\mathrm{M_{Jup}}$. Here, the uncertainty is propagated mainly from the uncertainty of $P_\mathrm{orb,min,3}$. This result means that the companion, if really exists, could be a brown dwarf or a low-mass star. Compared to the minimum-mass parameter value of $4~\mathrm{M_{Jup}}$, derived by \citet{Crouzet1}, the new value from this work is higher, and within the $1\sigma$ uncertainty, it excludes the possibility that the announced companion could have a planetary nature, see for example \citet{Rebolo1} and \citet{Burrows1}. 

\begin{table}
\centering
\caption{An overview of the fitted and derived parameters of the announced companion of XO-7b obtained from the quadratic RV model depicted in Fig. \ref{RVfit} (right-hand panel). For more details see the text in Sect. \ref{discuss}.}
\label{quadmodeltab}
\begin{tabular}{lll}
\hline
Parameter [unit] & Value & Comments\\
\hline
$O_\mathrm{SOPHIE}$ [$\mathrm{km~s^{-1}}$] & $-13.0823 \pm 0.0089$ & Fitted value\\
$O_\mathrm{MUSICOS}$ [$\mathrm{km~s^{-1}}$] & $-13.096 \pm 0.020$ & Fitted value\\
$d$ [$\mathrm{km~s^{-1}~d^{-1}}$] & $-0.000161 \pm 0.000018$ & Fitted value\\
$q$ [$\mathrm{km~s^{-1}~d^{-2}}$] & $0.000000172 \pm 0.000000036$ & Fitted value\\
$K_\mathrm{min,3}$ [$\mathrm{km~s^{-1}}$] & $0.1365 \pm 0.0005$ & Derived value\\
$P_\mathrm{orb,min,3}$ [d] & $7900 \pm 1660$ & Derived value\\
$(M_3 \sin i)_\mathrm{min}$ [$\mathrm{M_{Jup}}$] & $16.7 \pm 3.5$ & Derived value\\
\hline
\end{tabular}
\end{table}

Finally, we discuss the \textit{Gaia} Data Release No. 3 (DR3) Renormalized Unit Weight Error (RUWE, $\rho_\mathrm{R}$) value for XO-7 in terms of viability of the announced massive companion. The RUWE value allows distinguishing between 'good' and 'bad' astrometric solutions, where $\rho_\mathrm{R} \sim 1.4$ is the threshold between these solutions, and $\rho_\mathrm{R} \sim 1.0$ characterizes well-behaved single-star solutions \citep{Lindegren1, Lindegren2}. The \textit{Gaia} DR3 RUWE value for XO-7 is $\rho_\mathrm{R} = 1.116$ \citep{Gaia2}, which is relatively low and rather corresponds to a single-star solution. However, \citet{Belokurov1} conducted a detailed study of RUWE relation to unresolved binary systems in \textit{Gaia} Data Release No. 2 (DR2), selected a sample of 411 Jupiter hosts, and found that the peak of the RUWE distribution for the high-mass ($M_\mathrm{p} > 1~\mathrm{M_{Jup}}$) hot Jupiters ($P_\mathrm{orb} < 15$ d) is shifted to a value higher than $\rho_\mathrm{R} = 1$. In this group of Jupiters, the peak of the RUWE distribution is at $\rho_\mathrm{R} \sim 1.07$, and the authors concluded that this deviation is not induced by the known hot Jupiters, but the simplest interpretation is that these systems harbour an additional distant stellar or substellar component, which causes the photocentre to wobble. Although XO-7b does not fit $M_\mathrm{p} > 1~\mathrm{M_{Jup}}$, from this viewpoint, the \textit{Gaia} DR3 RUWE value for XO-7 seems to be not so low, and $\rho_\mathrm{R} = 1.116$ could indicate a distant companion present in the system. In order to shed more light on this, we converted the XO-7 RUWE value to the amplitude of the angular perturbation $\delta\theta$ \citep{Belokurov1}:

\begin{equation}
\label{Belokuroveq1}
\delta\theta \approx \sigma_\mathrm{AL}(G)~\sqrt{\rho_\mathrm{R}^2 -1},
\end{equation}

\noindent where the $\sigma_\mathrm{AL}$ as a function of source magnitude $G$ is presented in \citet{Lindegren1} as blue curve in their Fig. 9. Using $\sigma_\mathrm{AL} = 0.4$ mas in Eq. \ref{Belokuroveq1}, we obtained $\delta\theta \approx 0.198$ mas. Taking the distance dependence of the centroid wobble into account, the corresponding physical displacement in au is \citep{Belokurov1}:

\begin{equation}
\label{Belokuroveq2}
\delta a = \delta\theta~D,
\end{equation}

\noindent where $D$ is the distance to the source in kpc calculated as the inverse of parallax $p$ in mas. Using $p = 4.321 \pm 0.013$ mas for XO-7 from the \textit{Gaia} DR3 database \citep{Gaia2}, thus $D = 0.2314 \pm 0.0007$ kpc in Eq. \ref{Belokuroveq2}, we got $\delta a \approx 0.0458$ au. If we take into account that $P_\mathrm{orb,min,3} \gtrsim 7900 \pm 1660$ d and $(M_3 \sin i)_\mathrm{min} = 16.7 \pm 3.5~M_\mathrm{Jup}$, we can calculate the minimum semi-amplitude $a_\mathrm{min}$ of the star -- companion subsystem, based on Kepler's third law, as:

\begin{equation}
\label{Kepler3}
a_\mathrm{min} = [P_\mathrm{orb,min,3}^2 (M_\mathrm{s} + (M_3 \sin i)_\mathrm{min})]^{\frac{1}{3}}.
\end{equation}

\noindent In Eq. \ref{Kepler3} the orbital period is in years, the masses are in $\mathrm{M_\odot}$, $a_\mathrm{min}$ is expressed in au, we neglected XO-7b, and assumed a circular orbit. The result of this calculation, together with the uncertainty dictated mainly by the uncertainty of $P_\mathrm{orb,min,3}$, is $a_\mathrm{min} = 8.7 \pm 1.8$ au. If the companion is a brown dwarf or a low-mass star, we can assume that it is significantly fainter compared to XO-7, and therefore the perturbation should be due to the host star alone, and the photocenter should be the star itself. In this case, the minimum semi-amplitude of the photocentre perturbation $a_\mathrm{min,1}$ is:

\begin{equation}
\label{Kepler3-2}
a_\mathrm{min,1} = \frac{a_\mathrm{min} (M_3 \sin i)_\mathrm{min}}{M_\mathrm{s} + (M_3 \sin i)_\mathrm{min}}.
\end{equation}

\noindent Based on the Eq. \ref{Kepler3-2} we obtained $a_\mathrm{min,1} = 0.102 \pm 0.021$ au. This means that the amplitude of the photocentre perturbation should be $\delta a \gtrsim 0.204$ au, and not $\delta a \approx 0.0458$ au, as we calculated above. There are several possible explanations for this discrepancy. The first explanation is that there is no additional companion present in the system and the RUWE value ($\rho_\mathrm{R} = 1.116$) corresponds to a single-star solution. The second possibility is that the announced companion is a more massive stellar object and the system does not show any extra perturbation because the photocentre coincides with the centre of mass. However, this possibility is less probable, because a massive star should be visible in the spectra of XO-7. Finally, there is one more possibility, which we can mention in agreement with \citet{Belokurov1}, namely that the minimum orbital period of the companion ($P_\mathrm{orb,min,3} \gtrsim 7900 \pm 1660$ d) is much longer than the timebase of \textit{Gaia} individual data releases (e.g., 1038 days in the case of Early DR3, see \citet{Lindegren2}) and the photocentre perturbation will become quasi-linear and thus will be absorbed into the proper motion.

\section{Conclusions}
\label{concl}

We analyzed the XO-7 exoplanet system with the main scientific goal to follow-up the linear RV slope reported by the discoverers, and to put constraints on the orbital period of the announced wide-orbit massive companion of XO-7b. Furthermore, we aimed at refining the system parameters and we wanted to probe the TTVs of XO-7b in order to search for long-term dynamical signs of the companion of XO-7b in the O-C data of mid-transit times. To fulfill these aims we analyzed the 2-min integrated \textit{TESS} PDCSAP data from 8 sectors and performed a long-term RV monitoring of the planetary system. Moreover, in our analysis, we used the discovery RVs, as well. 

Our most interesting result is that the previously observed significant linear slope, $S_\mathrm{SOPHIE} = -0.1040 \pm 0.0065~\mathrm{km~s^{-1}~T_{obs}^{-1}}$, in the RVs was not confirmed with the follow-up RV data. We detected only a marginal linear slope, $S_\mathrm{MUSICOS} = 0.040 \pm 0.029~\mathrm{km~s^{-1}~T_{obs}^{-1}}$, in the newly obtained RV data, which has the opposite trend compared to the linear slope reported by the discoverers. It means that we did not find significant clues about the companion of XO-7b in the new RVs. We discussed three possibilities, what could be causing this discrepancy. We can conclude that if the announced companion really exists, the most convincing explanation is that both RV datasets were collected near its quadrature position. In this case, more RVs are needed to put better constraints on the orbital period of the companion. Based on the SOPHIE and MUSICOS RV observations of XO-7 we were able to estimate its lower limit, which is $P_\mathrm{orb,min,3} \gtrsim 7900 \pm 1660$ d. We did not find significant evidence of the companion of XO-7b in the O-C dataset of mid-transit times. We can again conclude that if the announced companion really exists, this is in agreement with previous results that distant companions of exoplanets are only known by RV solutions.

Furthermore, we used the quadratic RV solution of the merged SOPHIE and MUSICOS data to infer and discuss some preliminary physical parameters of the announced companion of XO-7b, mainly its mass. From this viewpoint, we can conclude that if the announced companion really exists, it could be a brown dwarf or a low-mass star with a 'minimum' minimum mass of $(M_3 \sin i)_\mathrm{min} = 16.7 \pm 3.5~\mathrm{M_{Jup}}$. We also discussed the \textit{Gaia} DR3 RUWE value for XO-7 in terms of viability of the companion. On the one hand, we can conclude that the RUWE value, $\rho_\mathrm{R} = 1.116$, rather corresponds to a single-star solution. On the other hand, there is a possibility that the astrometric perturbation effect of the companion is absorbed into the proper motion of the host star due to the orbital period, which is much longer than the timebase of \textit{Gaia} individual data releases.   

Finally, we obtained precise orbital and planetary parameters of the system via a combination of the \textit{TESS} data and RV observations of XO-7. Precise parameters of XO-7b, mainly the planet's mass and radius, will be important for further analyses. Mass determination can help constrain the internal structure of the planet, and break degeneracies in atmospheric characterization follow-up studies. Precise planet radius allows the planetary composition to be estimated. We also derived the fundamental parameters of the planet's host star. These parameter values are in a $3\sigma$ agreement to those presented by the discoverers, except for the projected rotational velocity, which we can explain with different spectral resolutions of the spectrographs.


\section*{Acknowledgements}

The authors thank P. Sivani\v{c} for the observations and technical assistance. This work was supported by the Hungarian National Research, Development and Innovation Office (NKFIH) grant K-125015, the PRODEX Experiment Agreement No. 4000137122 between the ELTE University and the European Space Agency (ESA-D/SCI-LE-2021-0025), the City of Szombathely under agreement No. 67.177-21/2016, and by the VEGA grant of the Slovak Academy of Sciences No. 2/0031/22. TP acknowledges support from the Slovak Research and Development Agency -- contract No. APVV-20-0148. This paper includes data collected with the TESS mission, obtained from the MAST data archive at the Space Telescope Science Institute (STScI). Funding for the TESS mission is provided by the NASA Explorer Program. STScI is operated by the Association of Universities for Research in Astronomy, Inc., under NASA contract NAS 5-26555. This work has made use of data from the European Space Agency (ESA) mission Gaia (\url{https://www.cosmos.esa.int/gaia}), processed by the Gaia Data Processing and Analysis Consortium (DPAC, \url{https://www.cosmos.esa.int/web/gaia/dpac/consortium}). Funding for the DPAC has been provided by national institutions, in particular, the institutions participating in the Gaia Multilateral Agreement.

\section*{Data availability}

The data underlying this article will be shared on reasonable request to the corresponding author.




\bibliographystyle{mnras}
\bibliography{Yourfile} 






\bsp	
\label{lastpage}
\end{document}